\documentclass{aa}
\usepackage{graphicx}
\usepackage{txfonts}
\usepackage{natbib}
\bibpunct{(}{)}{;}{a}{}{,}

\newcommand{\psr}{PSR~J1518$+$4904\,}
\newcommand{\omdot}{$\dot\omega$\,}
\newcommand{\omdoteq}{\dot\omega}
\newcommand{\pbdot}{$\dot {P}_\mathrm{b}$\,}

\newcommand{\xdot}{$\dot x$\,}
\newcommand{\xdoteq}{\dot x}
\newcommand{\msun}{$M_\odot$\,}
\newcommand{\tempo}{\textsc{tempo2}}

  \title{Multi-telescope timing of \psr}

  \author{G.\ H.\ Janssen\inst{1} 
    \and B.\ W.\ Stappers\inst{2,1,3}
    \and M. Kramer\inst{3}
    \and D.\ J.\ Nice\inst{4}
    \and A. Jessner\inst{5}
    \and I. Cognard \inst{6}
    \and M.\ B.\ Purver\inst{3}}

  \institute{Astronomical Institute ``Anton Pannekoek'', University of
  Amsterdam, Kruislaan 403, 1098 SJ Amsterdam, The Netherlands;\\  \email{gemma@science.uva.nl}
  \and Stichting ASTRON, Postbus 2, 7990 AA Dwingeloo, The
  Netherlands;  \email{Ben.Stappers@manchester.ac.uk} 
  \and University of Manchester, Jodrell Bank Observatory,
  Macclesfield Cheshire, SK11 9DL, UK
  \and Physics Department, Bryn Mawr College, Bryn Mawr, PA 19010, USA
  \and Max-Planck-Institut f\"ur Radioastronomie, Auf dem H\"ugel 69, 53121 Bonn, Germany
  \and Laboratoire de Physique et Chimie de l'Environnement, CNRS, 3A
  Avenue de la Recherche Scientifique, F-45071 Orl\'eans, Cedex 2,
  France}

  \date{Received/Accepted}  

\begin{document}

\abstract
{\psr is one of only 9 known double neutron star systems.
  These systems are highly valuable for measuring
  the masses of neutron stars, measuring the effects of gravity,
  and testing gravitational theories.  }
{We determine an improved timing solution for a mildly relativistic
  double neutron star system, combining data from multiple
  telescopes. We set better constraints on relativistic parameters
  and the separate masses of the system, and discuss the evolution of
  \psr in the context of other double neutron star systems.}
{\psr has been regularly observed for more than 10 years by the
  European Pulsar Timing Array (EPTA) network using the Westerbork, Jodrell Bank,
  Effelsberg and Nan\c cay radio telescopes. The data were
  analysed using the updated timing software \tempo.}
{We have improved the timing solution for this double neutron star
  system. The periastron advance has been refined and a significant
  detection of proper motion is presented. It is not likely that more
  post-Keplerian parameters, with which the individual neutron star
  masses and the inclination angle of the system can be determined
  separately, can be measured in the near future. }
{Using a combination of the high-quality data sets present in the EPTA
  collaboration, extended with the original GBT data, we have
  constrained the masses in the system to $m_\mathrm{p}<1.17$~\msun
  and $m_\mathrm{c}>1.55$~\msun ($95.4\% $ confidence), and the
  inclination angle of the orbit to be less than $47$ degrees
  (99\%). From this we derive that the pulsar in this system possibly
  has one of the lowest neutron star masses measured to date. From
  evolutionary considerations it seems likely that the companion star,
  despite its high mass, was formed in an electron-capture supernova.
  }


\maketitle

\section{Introduction}

\psr was discovered in the Green Bank Northern Sky Survey
\citep{snt97}. It is one of only 9 double neutron star systems (DNSs)
known.  The pulsar orbits its companion neutron star in 8.63~days, and
its 40~ms spin period, together with the derived low surface magnetic
field, is typical of a mildly recycled pulsar.  \cite{nst96} first
described the system and have already shown that the space velocity is
probably quite low. They also measured the periastron advance
from which the total system mass was estimated.  \cite{tc99}, using a
Bayesian analysis with no constraints on the orientation of the orbit,
found the masses of the pulsar and its companion to be $m_\mathrm{p}=
1.56^{+0.20}_{-1.20}$ \msun and $m_\mathrm{c}= 1.05^{+1.21}_{-0.14}$
\msun ($95\% $\,confidence).  An improved timing solution was
presented by \cite{hlk+04}, although not discussed in detail.

DNSs are presently the best available tool for testing strong-field
gravity effects.  The \psr\ orbit is only mildly relativistic, making
it of limited use for tests of gravitational theories.  However,
any constraints on its post-Keplerian (PK) parameters, the
inclination of its orbit, or the masses of the pulsar and its
companion are highly valuable for studies of the evolution of these
systems (e.g. \citealt{plp+04,kbk+07}).

Like PSR~J1811$-$1736 \citep{cks+07}, \psr\ has a rather wide orbit. This could
indicate that the evolution into a DNS system has
been slightly different than most other systems, which generally have
tight orbits with periods of several hours. However, \psr\ does follow
the recently discovered spin period-eccentricity relation
\citep{fkl+05,dpp05}, suggesting that the evolution cannot be too 
different from the tight DNSs. 

\psr has been observed regularly using the four 100~m class radio
telescopes in Europe. This has allowed us to select high-quality data
resulting in the best possible analysis of the system parameters to
date. For completeness, we have also included data from telescopes at
Green Bank as presented in \cite{nst96, nts99}.  We present the new
timing solution and discuss the limits our solution puts on the
inclination and the masses of the neutron stars in the system.  We
discuss the future prospects of detecting multiple PK parameters, and
we related our results to the evolution of DNSs.

\section{Observations and data analysis}

\begin{table*}
  \centering
  \caption{Individual data sets. 
    \label{tab:dataset}}
  \begin{tabular}{lcccccc}
    \\*[-4ex]
    \hline \hline \\*[-2ex]
    & Effelsberg$^b$ & Jodrell Bank & Nan\c cay$^b$ & Westerbork &
    Green Bank$^b$ & All combined \\
   \hline \noalign{\smallskip}
    Number of TOAs & 71 & 292 & 145 & 126 & 382 & 1016 \\
    Time span (MJD) & 52481 -- 54166 & 49797 -- 53925 & 53307 -- 54200 &
    51389 -- 54337 & 49670 -- 52895 & 49670 -- 54337 \\ 
    R.M.S. individual data set$^a$ ($\mu$s)& 3.3 & 10.7 & 3.3 & 5.2 & 8.2 & 6.0 \\
    Observed Frequencies (MHz) & 1400 & 400, 600, 1400 & 1400 & 840,
    1380, 2300 & 350, 370, 575, 800 & \\
  \hline 
  \hline \\*[-2ex]
  \end{tabular}
  The best timing solution acquired from all data sets is
    presented in Table ~\ref{tab:jwnge-2}. $(a)$ R.M.S. given require
    reduced $\chi^2 = 1$ for each individual data set. $(b)$ R.M.S. is the result
    of a fit with proper motion (Green Bank, Effelsberg, Nan\c cay)
    and dispersion measure (Effelsberg, Nan\c cay) fixed.
\end{table*}

This research is part of the EPTA network.  The extensive collection
of observations from Westerbork, Jodrell Bank, Effelsberg and Nan\c
cay observatories has enabled us to select the best combinations of
data sets with the best characteristics for various purposes.
For example, high quality TOAs and long time spans permit 
better measurement of  astrometric parameters, while using multiple
frequencies allows us to precisely measure the dispersion measure (DM) and
monitor possible DM variations. Individual properties of the data sets
are presented in Table \ref{tab:dataset}, and a description of the
observing systems is given below.

\subsection{Westerbork}

\begin{figure}
  \centering
  \includegraphics[width=6.5cm, angle=270]{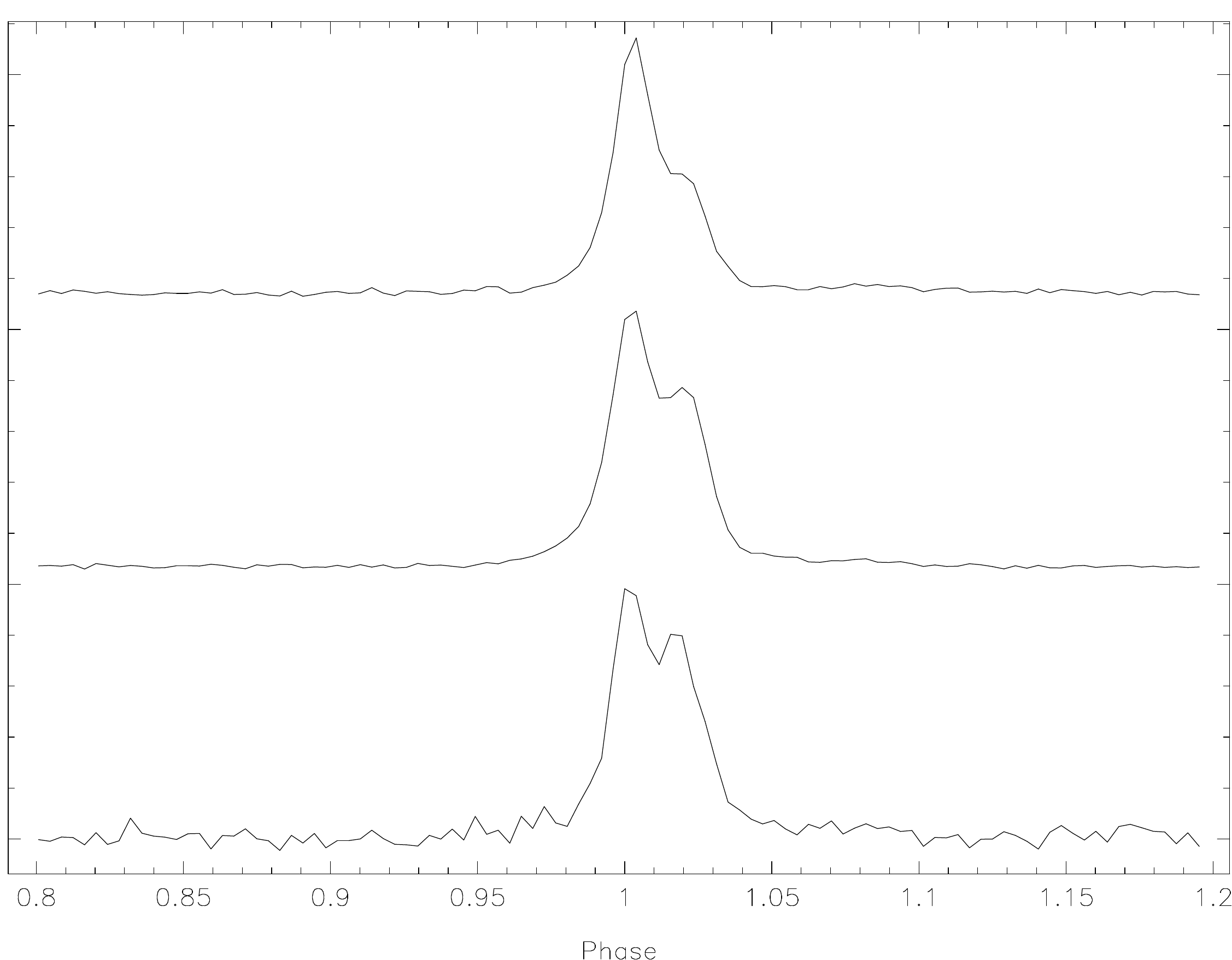}
  \caption{Standard profiles for the three different frequencies used
  at Westerbork. From top to bottom: 840 MHz, 1380 MHz, 2300 MHz. The
  effective time resolution ($102.4\ \mu$s) is similar to the size of
  one bin ($160\ \mu$s).
  \label{fig:stds}}
\end{figure}

\psr has been observed approximately monthly since 1999 at the Westerbork
Synthesis Radio Telescope (WSRT) with the Pulsar Machine (PuMa;
\citealt{vkh+02}), at frequencies centred at 840 and 1380 MHz, and
since 2007 also occasionally at 2300 MHz.  The sampling time for most
observations was 102.4~$\mu$s
and the bandwidth used was 8$\times$10~MHz where each 10 MHz band was
split into 64 channels. The 1380 MHz data taken after September 2006
used 80 MHz of bandwidth spread in 8 steps of 10 MHz over a range of
160 MHz. 
The data were dedispersed and folded off-line, and then integrated over
frequency and time over the whole observation duration to get a single
profile for each observation.  Each profile was cross-correlated with
a standard profile (Fig. \ref{fig:stds}), obtained from the summation
of high signal-to-noise (S/N) profiles, to calculate a time of arrival
(TOA) for each observation. These were referred to local time using
time stamps from a H-maser at WSRT.  The TOAs were converted to UTC
using global positioning system (GPS) maser offset values measured at 
the observatory.
For this research, we only used observations with TOA errors $<
15\,\mu s$, with an average around 8\,$\mu$s.

\subsection{Jodrell Bank}

At Jodrell Bank, \psr was observed since its discovery
approximately every two weeks using the 76-m Lovell telescope.
Observations were made at centre frequencies around 410, 610 and 1400 MHz.
All receivers were cryogenically cooled, providing left-hand and
right-hand circularly polarized signals. These signals were fed into an
analogue filterbank system where the polarisations were detected,
filtered, digitised at appropriate sampling intervals and incoherently
dedispersed in hardware. The sampling time was chosen to match the
dispersion smearing across one filterbank channel. At 410 MHz, a $2\times
32 \times 0.0312$ MHz system was used, at 610 MHz we employed a
$2\times 6 \times0.1250$ MHz filterbank, while the vast majority of our
data were recorded with a $2\times 32 \times 1$ MHz system at 1400 MHz.

The resulting dedispersed timeseries were folded on-line with the
topocentric pulsar period and finally written to disc. In the off-line
reduction, the two polarisations were summed to form total-intensity
profiles. A standard pulse template was used at each frequency to
determine the TOA. During this process, TOAs were referred to the local
H-maser time-standard and already corrected to UTC using information
obtained via the GPS. In our analysis, we rejected TOAs with estimated
uncertainties exceeding 40\,$\mu$s (or $60\,\mu$s for data prior to Jan 2000,
respectively). With an average TOA error for the final data set of
18\,$\mu$s, the precision of this incoherently dedispersed data set is
modest, but it provides the longest time baseline in our analysis. The
Jodrell Bank data were therefore used as a reference to calculate
observatory-related offsets between the other data sets used.

\subsection{Effelsberg}

Observations in Effelsberg were made at least once a month
at 1410 MHz and at irregular intervals at 850 MHz. At 1410 MHz a
tunable HEMT receiver was used achieving typical system temperatures
of 30-40 K depending on weather conditions and telescope elevation. 

At the highest frequency a total bandwith of 55 MHz
was available from the coherently dedispersing backend. For each
polarisation the band was split into four subbands which themselves
were subdivided into eight digitally sampled channels.  Each of these
32 bands per polarisation was coherently dedispersed by programmable
digital filters and then recombined with the approriate channel
dispersion delay \citep{bdz+97} and synchronously summed up with
the current topocentric period.

A TOA was calculated for each average profile obtained during a 5-10
minute observation. During this process, the observed time-stamped
profile was compared to a synthetic template, which was
constructed out of 3 Gaussian components fitted to a high
S/N standard profile \citep{kxl+98,kll+99}. This
template matching was done by a least-squares fitting of the
Fourier-transformed data \citep{tay92}. Using the measured time delay
between the actual profile and the template, the accurate time stamp
of the data provided by a local H-maser and corrected off-line to UTC
using recorded information from GPS satellites, the final TOA was
obtained.

Originally, multiple TOAs were calculated from one observation,
but better results were acquired from calculating one TOA per
observation. We have used data obtained since August 2002, having TOA
errors smaller than 20 $\mu$s, on average 5$\mu$s.

\subsection{Nan\c cay}

The pulsar has been observed weekly in Nan\c cay since late
2004. Equivalent to a 93-m dish, the Nan\c cay Radio Telescope and the
BON (Berkeley-Orl\'eans-Nan\c cay) coherent dedispersor were used for
typical integration times of 45 min. Coherent dedispersion of a 64 MHz
band centred on 1398 MHz is carried out on sixteen 4 MHz channels using a
PC-cluster. The Nan\c cay data are recorded on a UTC(GPS) time scale
built at the analogue to digital converter by a Thunderbolt receiver (Trimble
Inc.). Differences between UTC and UTC(GPS) are less than 10~ns at
maximum, and therefore no laboratory clock corrections are
needed. One TOA was calculated from a cross-correlation with a
pulse template for each observation of $\sim$~45 min. 
Except for 24 low S/N profiles that were excluded from the used
sample, all  
TOAs have errors less than 15~$\mu$s, but generally around 5~$\mu$s. 
Since the data span of Nan\c cay Observatory covers only a few
years, the proper motion of \psr was fixed in determining
the best timing solution for this data set alone.

\subsection{Green Bank}

\psr was observed at Green Bank using the 140 Foot (43
m) Telescope and the 100 m Green Bank Telescope (GBT).  Observations
with
the 140 Foot Telescope were made every few weeks from November 1994 to
July 1999 at 575 and 800 MHz.  Intensive campaigns, giving full
coverage of the orbit over several days, were made with the 140 Foot
Telescope at 370 MHz in March/April 1995, August/September 1995, May
1998, and December 1998.  An intensive campaign was made using the
GBT at 370 MHz in August/September 2003.

The same data acquisition system and nearly identical observational
procedures were employed at the two telescopes.  Details are given in
\cite{nst96}.  Briefly, data were collected by the Spectral
Processor, a Fourier transform spectrometer, which synthesized 512
spectral channels across a passband of 40 MHz and folded the power
measurements in each channel at the topocentric pulse period to produce pulse
profiles.  Observation times ranged from a few
minutes to several hours on any given day.  During the course of an
observation, data were integrated over intervals of 2 minutes (140
Foot Telescope) or 0.5 minutes (GBT).  Data were
separately collected in two polarizations.  Off-line, each integration
was processed by summing polarizations, dedispersing and summing
spectral channels to produce a single pulse profile, and deriving a
TOA using conventional procedures.  Observations of up to one hour
were averaged together by computing TOAs for individual integrations
and averaging them into a single effective TOA for the hour-long span.
Time stamps were provided by a local H-maser and corrected off-line to
UTC using recorded information from GPS satellites.
Arbitrary offsets were allowed in the timing solution between
sets of TOAs collected at different frequencies, and between two
sets of GBT TOAs collected with different receiver polarization
settings.

\subsection{Combination of data sets}

Combining data sets obtained by different observatories will have many
advantages.  Using a longer time span sampled with more data is
usually better for timing in general.  The four telescopes now in use
for EPTA timing purposes all contribute differently to the total
picture of high-precision timing.  The Westerbork, Effelsberg and
Nan\c cay data sets are of very high quality, and adding the Jodrell
Bank long-term data which has only slightly lower average R.M.S. provides
us with both a good coverage in time as well as in orbital phase for
this pulsar. The addition of the original GBT data further complements
our set of measurements of TOAs from the system.

When combining data from multiple telescopes, all with different
observing and operating systems, it is important to account for all
extra time corrections needed, apart from the usual correction from
arrival times at individual telescopes to arrival times in TAI at the
solar system barycentre\footnote{ftp://ssd.jpl.nasa.gov/pub/eph/export/DE405/de405iom.ps} (SSB).
Observatory-related time delays that are not accounted for
(i.e. unmodelled cable delays) as well as a different approach in
calculating TOAs (template differences) will result in a
time offset between sets of residuals from the different
telescopes. The new timing software package \tempo\ \citep{hem06} is
capable of fitting for constant time offsets, and these so-called
``jumps'', which are on the order of $0.1$ ms, have been used as extra
parameters in the timing model.  A more detailed study on this subject
will be presented in a forthcoming paper.

\section{Results}

\begin{table}
\caption{Timing solution using data from Jodrell Bank, Westerbork,
  Effelsberg, Nan\c cay and Green Bank. 
  \label{tab:jwnge-2}}
\begin{tabular}{ll}
\\[-4ex]
\hline\hline \\*[-2ex]
\multicolumn{2}{c}{Fit and data-set} \\
\hline  \\*[-2ex]
Pulsar name\dotfill & J1518+4904 \\ 
MJD range\dotfill & 49670.7---54337.7 \\ 
Number of TOAs\dotfill & 1016 \\
Epoch (MJD)\dotfill & 52000 \\ 
R.M.S. timing residual ($\mu s$)\dotfill & 6.05 \\
 Weighted fit\dotfill &  Y \\ 
Reduced $\chi^2$ value \dotfill & 1.117 \\
\hline \\*[-2ex]
\multicolumn{2}{c}{Measured Quantities} \\ 
\hline \\*[-2ex]
Right ascension, $\alpha$ (J2000)\dotfill &  15$^\mathrm{h}$18$^\mathrm{m}$16\fs799084(16) \\ 
Declination, $\delta$ (J2000)\dotfill & +49\degr04\arcmin34\farcs25119(16) \\ 
Pulse frequency, $\nu$ (s$^{-1}$)\dotfill & 24.4289793809236(3) \\ 
First derivative of pulse frequency, $\dot{\nu}$ (s$^{-2}$)\dotfill & $-$1.62263(12)$\times 10^{-17}$ \\ 
Dispersion measure, $DM$ (cm$^{-3}$pc)\dotfill & 11.61139(8) \\ 
Proper motion in RA, $\mu_\mathrm{\alpha}$ (mas\,yr$^{-1}$)\dotfill & $-$0.67(4) \\ 
Proper motion in DEC, $\mu_\mathrm{\delta}$ (mas\,yr$^{-1}$)\dotfill & $-$8.53(4) \\ 
Orbital period, $P_\mathrm{b}$ (d)\dotfill & 8.6340050964(11) \\ 
Epoch of periastron, $T_0$ (MJD)\dotfill & 52857.71084163(17) \\ 
Projected semi-major axis of orbit, $x$ (lt-s)\dotfill & 20.0440029(4) \\ 
Longitude of periastron, $\omega_\mathrm{0}$ (deg)\dotfill & 342.554394(7) \\ 
Orbital eccentricity, $e$\dotfill & 0.24948451(3) \\ 
First derivative of orbital period, $\dot{P}_\mathrm{b}$\dotfill & 2.4(22)$\times 10^{-13}$ \\ 
First derivative of $x$, $\dot{x}$\dotfill & $-$1.1(3)$\times 10^{-14}$ \\ 
Periastron advance, $\dot{\omega}$ (deg\,yr$^{-1}$)\dotfill & 0.0113725(19) \\ 
\hline  \\*[-2ex]
    \noalign{\smallskip}
\multicolumn{2}{c}{Derived Quantities} \\
\hline \\*[-2ex]
$\log_{10}$(Characteristic age, yr) \dotfill & 10.38 \\
$\log_{10}$(Surface magnetic field strength, G) \dotfill & 9.03 \\
Mass function (\msun)\dotfill & 0.115988 \\
Total mass, $M_\mathrm{T}$ (\msun)\dotfill & 2.7183(7)\\
Total proper motion, $\mu_\mathrm{T}$ (mas\,yr$^{-1}$)\dotfill & 8.55(4)\\
DM distance (pc)\dotfill & $625^{+90}_{-83}$ \\
\hline \\*[-2ex]
    \noalign{\smallskip}
\multicolumn{2}{c}{Assumptions} \\
\hline \\*[-2ex]
Clock correction procedure\dotfill & TT(TAI) \\
Solar system ephemeris model\dotfill & DE405 \\
Binary model\dotfill & DD \\
Model version number\dotfill & 5.00 \\ 
\hline
\hline \\*[-2ex]
\end{tabular}
Figures in parentheses are the nominal 1$\sigma$ \textsc{tempo2}
uncertainties in the least-significant digits quoted. 
  All time-dependent variables refer to the same epoch.
  The DM distance is estimated from the \cite{cl02} model.
\end{table}

\begin{figure}
  \centering
  \includegraphics[width=6cm, angle=270]{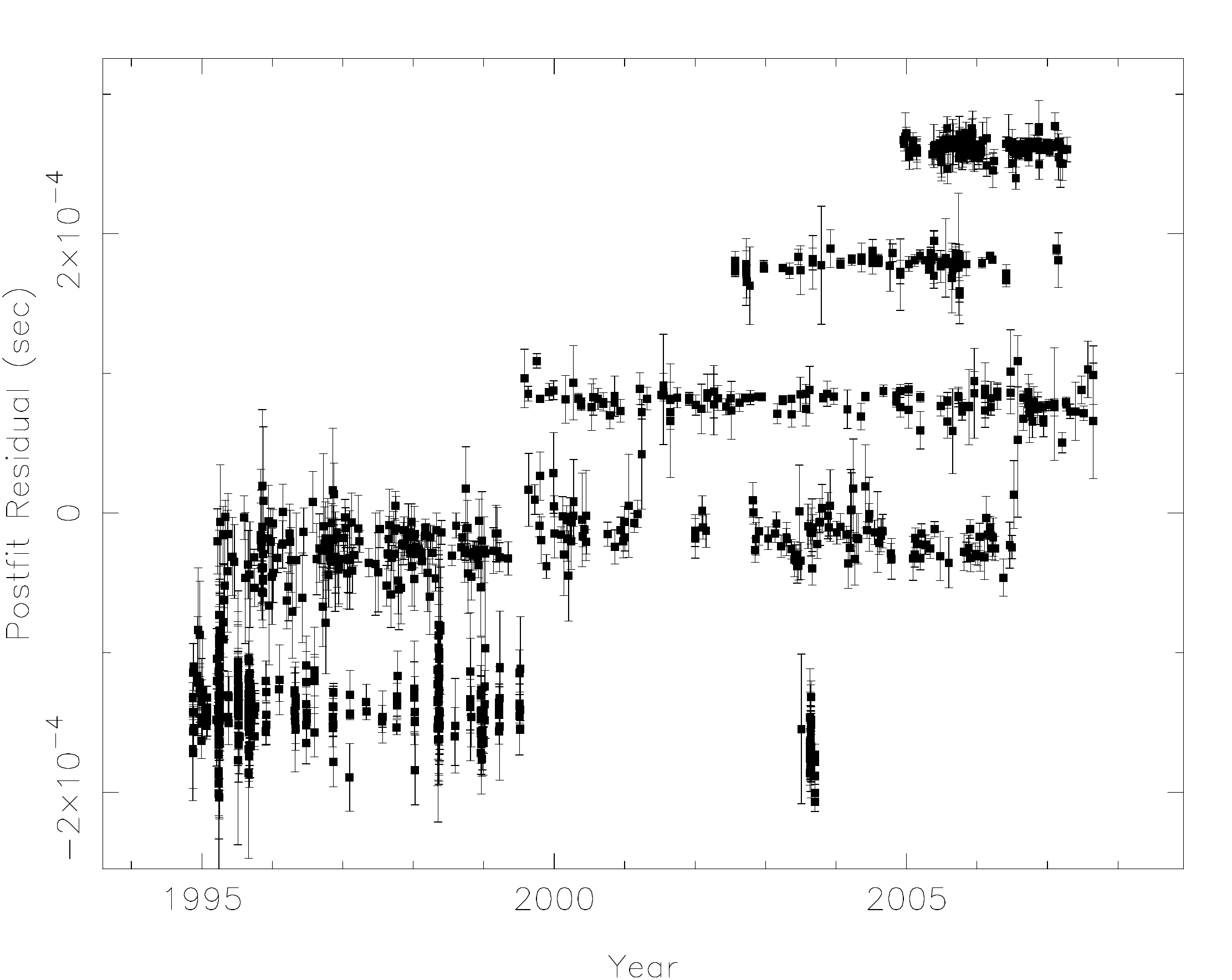}
  \includegraphics[width=6cm, angle=270]{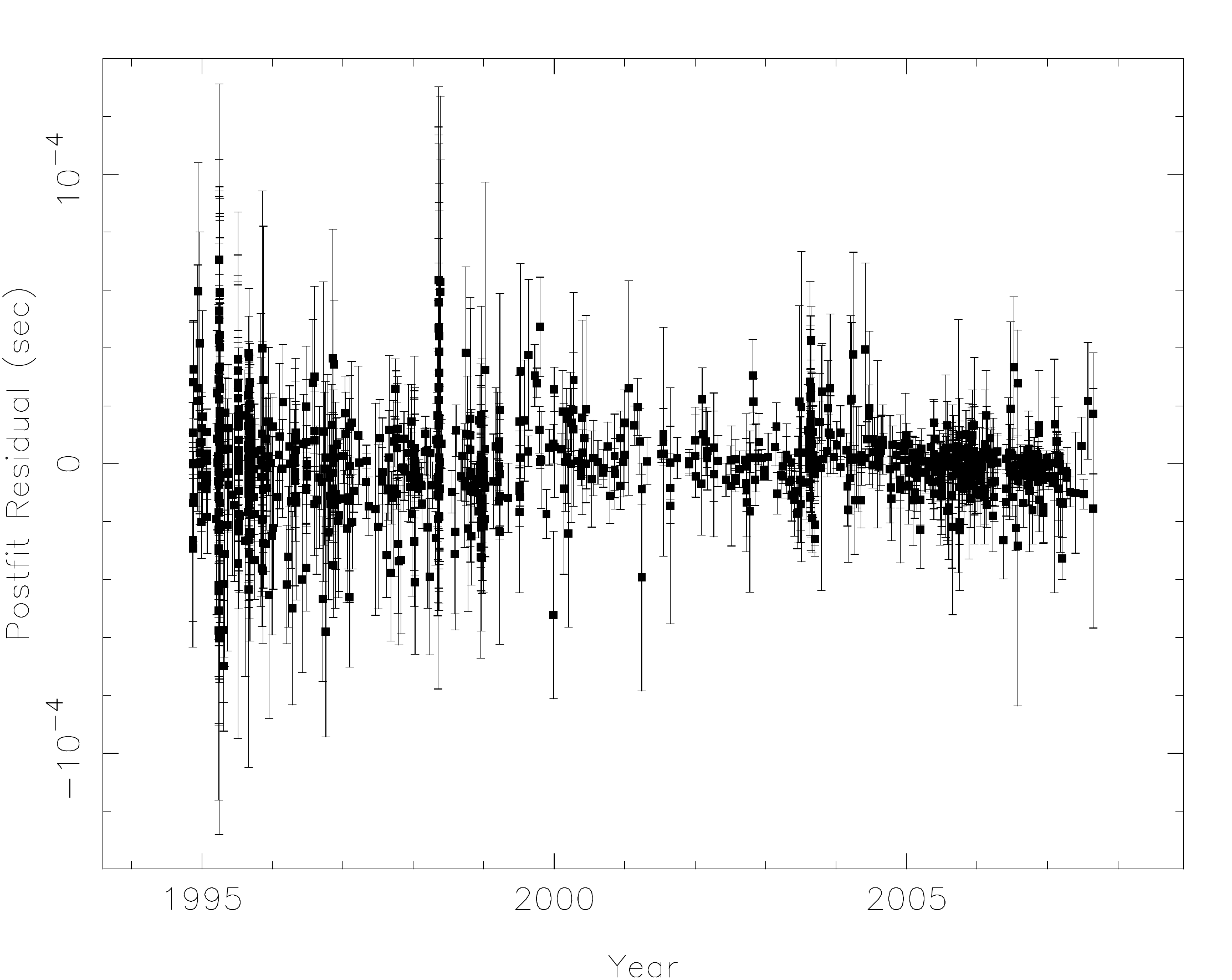}
  \includegraphics[width=6cm, angle=270]{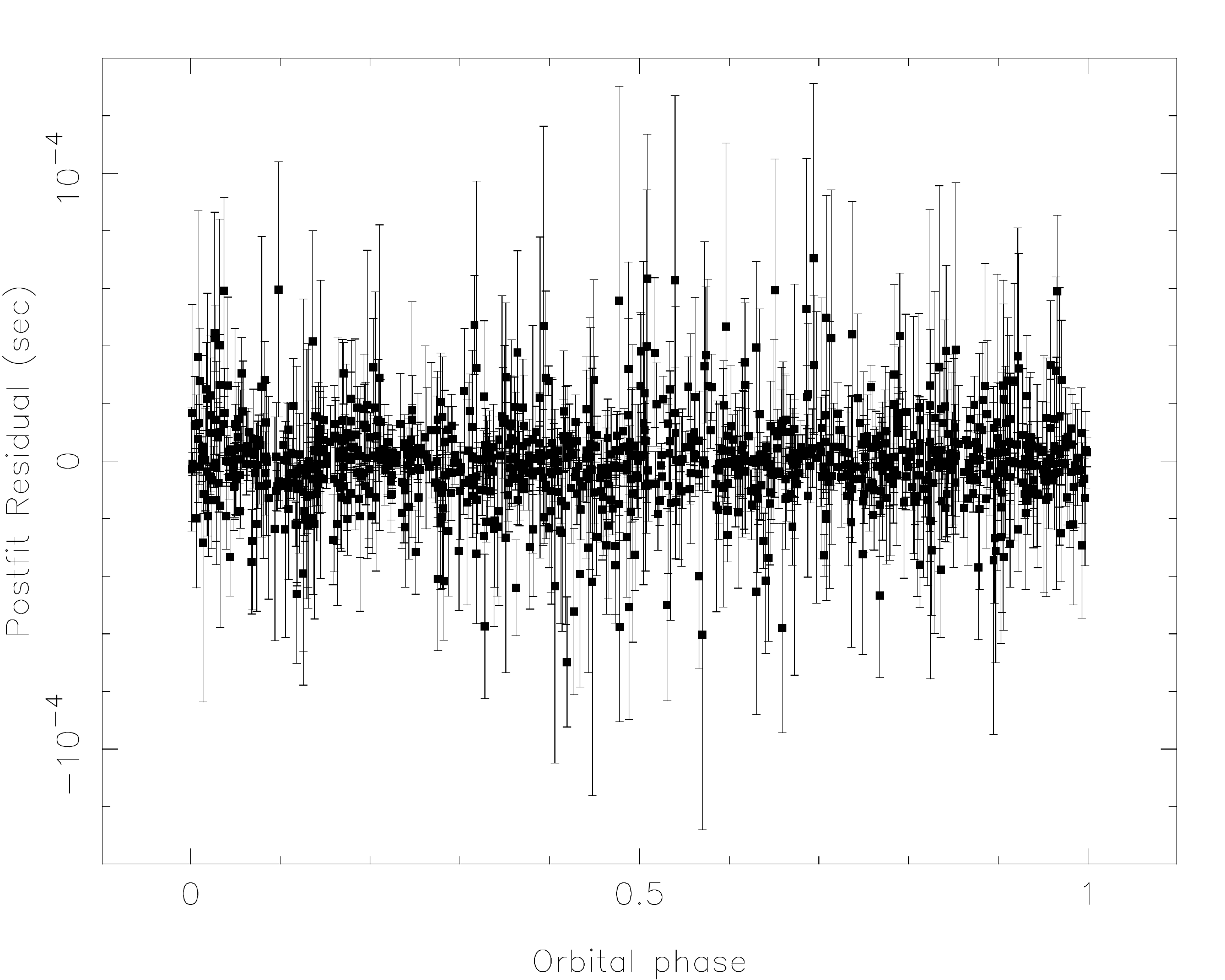}
  \caption{Best fit timing residuals including all timing data
  available. The top panel shows the different data sets, from bottom
  to top: GBT, Jodrell Bank, Westerbork, Effelsberg and Nan\c cay. The
  offsets have been adapted manually to show the data sets
  separately. This plot was generated from the best timing solution,
  which is shown in the middle panel as residuals vs time, and
  residuals vs orbital phase in the bottom panel. This timing solution has an rms
  of 6 $\mu$s and the parameters are listed in Table \ref{tab:jwnge-2}.
  \label{fig:jwnge-2}}
\end{figure}

The best overall timing solution for \psr, obtained by combining data
from Jodrell Bank, Westerbork, Effelsberg, Nan\c cay and Green Bank,
is presented in Fig. \ref{fig:jwnge-2} and Tables \ref{tab:dataset}
and \ref{tab:jwnge-2}.  Before combining the data sets from the
different observatories, the errors on the TOAs were scaled with a
constant factor to have each individual data set return a reduced
$\chi^2\approx1$.

Having multiple sets of TOAs allows us to select different subsets of
TOAs for refining different parameters.  When measuring the
astrometric and rotational parameters like position, proper motion and
spin frequencies, it is best to use a combination covering a data span
as long as possible, in our case using all available data sets.  Even
though the standard deviation of the residuals from zero was, on
average, somewhat larger, this yields the best possible determination
of these parameters.

For short-period binaries, it can sometimes be better to select only
the data set(s) with the highest precision TOAs to determine the binary
parameters more accurately. We have tried several combinations of the
best TOAs, i.e. using Nan\c cay and Westerbork, or Nan\c cay,
Westerbork and Effelsberg. 
However, we found the best parameter measurements are derived from
timing solutions which combine all data sets, not just those with
the most precise TOAs. The total data set has broader orbital
coverage compared to any combination of subsets having the most precise TOAs,
and also covers a longer time span, improving the measurement
precision of parameters that undergo secular changes.
The timing parameters derived from various combinations are all in
agreement with each other, showing that there are no large systematic
effects in the separate data sets.

\subsection{Dispersion measure variations}

Although the majority of all observations were done at frequencies around 1400
MHz, Westerbork, Jodrell Bank and Green Bank have also observed \psr 
at other frequencies, see Table \ref{tab:dataset}, allowing us, in
principle, to detect DM variations.
We applied the ``stridefit'' plugin of \tempo \, which fits small
segments of time for DM while keeping all other parameters fixed.
Because the observations at additional frequencies are not distributed
evenly across the full time span, it was not possible to apply this
method continuously.  However, there is no evidence of large
variations or trends present in the dispersion measure.

In a similar way as explained above,
we have tried to improve our timing solution by fixing the best DM
found from the solution using all available TOAs, and fitting again
using only the higher precision 1400 MHz data from Jodrell Bank, Westerbork, Nan\c cay and
Effelsberg. Again, having the data set as extended as possible resulted
in the better determination of the parameters.

\subsection{Parallax}

Even though the DM is quite small, the error in the DM-derived
distance is still dominating the significance of any velocity or
orientiation measurement.
For a relatively nearby system like \psr, it is worthwhile investigating
the possibility of detecting a parallax signature in the timing
residuals to be able to acquire an independent, and perhaps even better,
distance estimate.

The expected amplitude in the timing residuals due to parallax (from
\citealt{lk05}, \S 8.2.5) is 
$l^2 \cos \beta/2 c d~\sim~0.9 \mu s$,
where $l$ is the distance between the Earth and the SSB, $\beta$ the
ecliptic latitude of the pulsar, $c$ the speed of light and $d$ the
distance to the pulsar.
Unless the TOA measurements improve significantly, we cannot expect
to measure the parallax for at least another 600 years.

\subsection{Proper motion}

As part of the astrometric fit, we have made a significant detection
of the proper motion of this system, in right ascension $\mu_\mathrm{\alpha}$ =
$-$0.67(4), and in declination $\mu_\mathrm{\delta}$ = $-$8.53(4) mas\,yr$^{-1}$. The
total proper motion is $\mu_\mathrm{T} $ = 8.55(7) mas\,yr$^{-1}$ which, using the
DM-derived\footnote{The DM distance is estimated from the \cite{cl02}
model.} distance of $625^{+90}_{-83}$ pc, corresponds to a transverse
velocity of just 25(4) km\,s$^{-1}$.

\subsection{Shklovskii effect}

The transverse motion of the system results in an increasing projected
distance of the pulsar to the solar system barycentre, which affects
any observed change in periodicities in the system \citep{shk70}.

For this pulsar, with a period of $40.93$ ms, a total proper motion
measured to be $8.55$ mas yr$^{-1}$ and distance estimated at 625 pc,
the Shklovskii term in the period derivative will be as large as
$20\%$ of the observed $\dot P$. 
However, as will be shown in \S\ref{section:pbdot}, this effect can be cancelled
out by the acceleration towards the Galactic disc. The observed period
derivative of the pulsar is therefore likely to be close to the
intrinsic value. The estimate of the characteristic age $\tau_\mathrm{c}$ and
the magnetic field will not deviate significantly from the \tempo\,
calculated values as presented in Table \ref{tab:jwnge-2}.

\subsection{Secular perturbation of x and\ \omdot}\label{section:xdot}

Because of the significant proper motion measurement, it is
interesting to investigate any geometric effects that are likely to
result from changes in the line of sight towards the system. For
example, the varying projection of the orbit as the system moves
through space can result in an apparent change in the semimajor axis
of the orbit, \xdot, and a perturbation of the observed periastron
advance,\, $\delta$\omdot. Following \cite{kop96}, but writing in
terms of position angles, these can be expressed as:

\begin{equation}\label{eq:xdot}
\xdoteq = x\  \mu_\mathrm{T} \cot i \cos (\Theta_\mathrm{\mu}-\Omega),
\end{equation}

\begin{equation}\label{eq:omdot}
\delta \omdoteq\ = -\mu_\mathrm{T} \csc i \sin (\Theta_\mathrm{\mu}-\Omega),
\end{equation}
where $\Theta_\mathrm{\mu}$ and $\Omega$ represent the direction of
proper motion and the position angle of the ascending node
respectively (measured North through East).
When it can be shown that the proper motion is the dominant effect on
the observed \xdot, 
Eq.~\ref{eq:xdot} can be used to derive an upper limit on the orbital inclination:
\begin{equation}
\tan i < \mu_\mathrm{T} / \left( \frac{\xdoteq}{x} \right),
\end{equation}
where we have used $|\cos (\Theta_\mathrm{\mu}-\Omega) | \leq 1$.
In this system, the effect of gravitational wave (GW) emission on the
change of semi-major axis will be negligible, as it is expected to be
only a fraction of the GW-effect on \pbdot, which is already
undetectable (see \S \ref{section:pbdot}), so that we can apply the above formula.

We were able to significantly measure a change in
the projected semi-major axis: \xdot\, $ = -1.1(3)\times10^{-14}$.
Taken at face value, our measurement corresponds to an upper
limit on the inclination of $69$ degrees, assuming  $0^\circ<i<90^\circ$.
A complete analysis of the interpretation of \xdot, including all
orbital orientations and including the effects of its covariance with
the relativistic parameter $\gamma$, will be presented in
\S\ref{section:masses} below.

\subsection{Post-Keplerian parameters}

As mentioned above, the wide orbit of \psr\, means that it is unlikely
that another PK parameter, besides \omdot,
will be detected in the near future.  
However, considering that \pbdot\ is the PK parameter with the
strongest dependence on the total observing time span $T$ and 
$P_b$ (\citealt{dt92}, Table II), it will probably be measured 
sooner than any of the other PK parameters. 
Placing an exact value on the timescale it will take to measure any 
further PK parameters is difficult, due to the covariances 
with other parameters and the possibility that kinematic terms may come 
to dominate the timing solution.
In general, measuring any further PK parameters in this system will be 
challenging and will require observation durations that at least double 
the existing observation span of 12.8 years. 

We consider each of the PK parameters below.  For this system we
neglect any contributions to the PK parameters due to effects other
than general relativity (GR) or proper motion.  As spin-orbit coupling and
tidal effects are only expected to be of influence for a main-sequence
companion, and any significant third body in the system would have
been detected in the timing residuals, we consider this a valid
assumption.

\subsubsection{Advance of periastron: \omdot}

We measure a value of \omdot = 0.0113725(19) deg\,yr$^{-1}$ for the periastron
advance of the orbit. This agrees with, and improves upon the already
published values of 0.0111(2) deg\,yr$^{-1}$, 0.0113(1) deg\,yr$^{-1}$
\citep{nst96,nts99} and 0.01138(4) deg\,yr$^{-1}$ \citep{hlk+04}.

Our measured \omdot, combined with the mass function resulting from
the orbital parameters, gives a very precise total mass of the system
of $2.7183(7)$ \msun. This value is calculated assuming GR is the
correct theory of gravity and the only measurable cause of periastron
advance. 
As can be seen from Eq.\ref{eq:omdot},
and shown in more detail
in Fig.~\ref{fig:paramplot}, this is likely justified, as we find the
contribution to \omdot\ from the proper motion for the possible
geometric configurations of the system to be of the same order as the
measurement uncertainty.

\subsubsection{Shapiro delay parameters $r$ and $s$}\label{section:shapiro}

\begin{figure}
  \centering
  \includegraphics[width=7.5cm, angle=270]{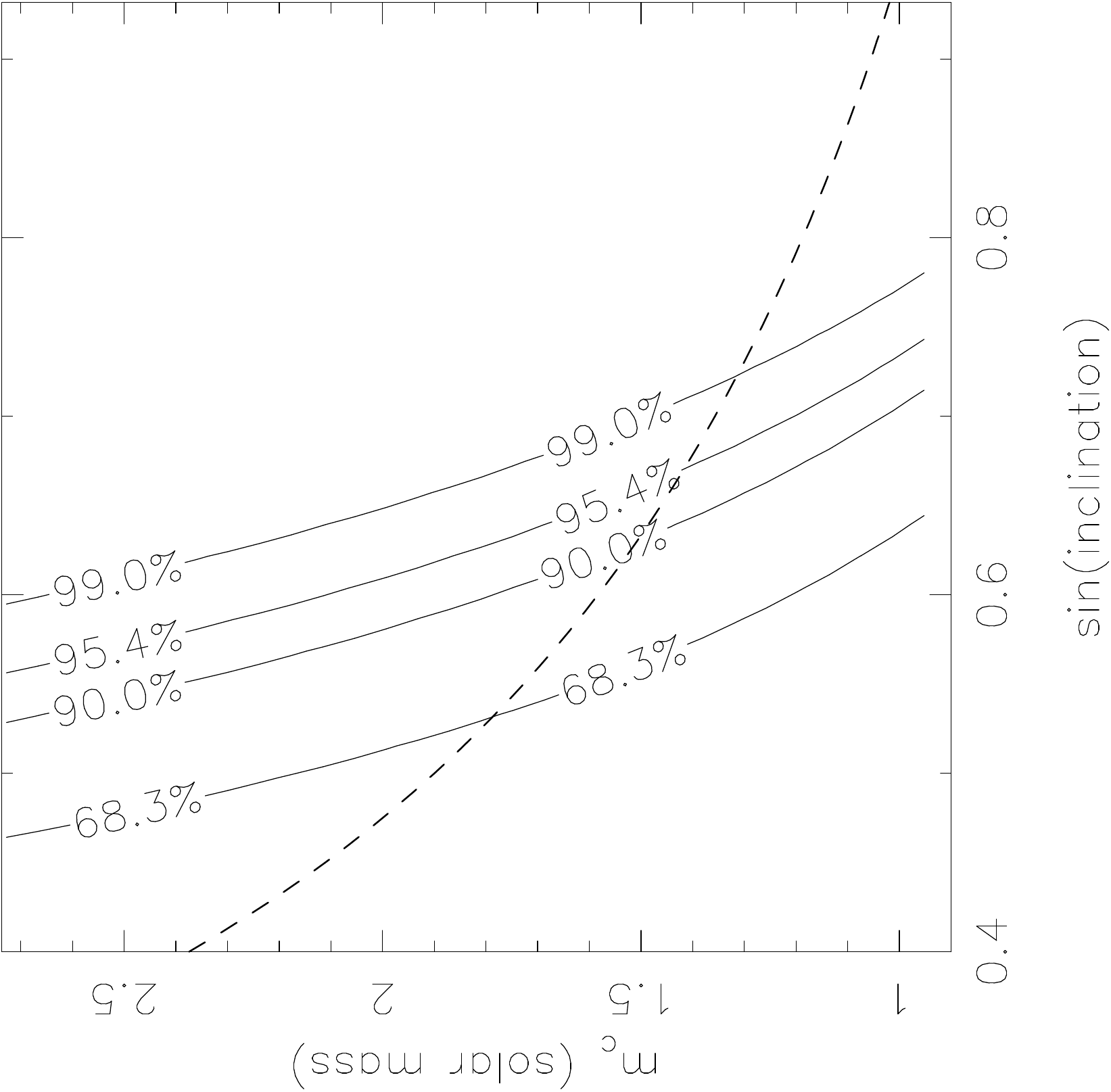}
  \caption{Chisq contour plot of $m_\mathrm{c}$-$\sin i$ parameter space. Lines
   with probabilities indicate regions with certain confidence
   levels. The dashed line represents the constraint on total mass
   resulting from the \omdot measurement. From the non-detection of
   shapiro delay we can set an upper limit to the inclination of the
   orbit: $i < 47$ degrees. This corresponds to a lower limit for the
   companion mass of 1.29 \msun.
  \label{fig:m2sininew}}
\end{figure}

For suitable inclination angles, Shapiro delay would be readily
detected in this system. However, we have not seen the signature of
Shapiro delay in the data, which indicates that the orbit has a low
inclination. We use this to set a limit on the inclination of the orbit
by generating a map of $\chi^2$ space
resulting from fitting the mass of the companion and inclination angle
in various combinations\footnote{This analysis used the \tempo \, plugin
{\it m2sini}. For explanation of this plugin, see \cite{ojh+06}.}.  The
resulting probability contours are shown in Fig. \ref{fig:m2sininew},
along with the constraint on the total mass from our \omdot\
measurement.  In our case the contour plot results in upper limits
only and there is a 99\% chance that $\sin~i \le 0.73$, corresponding
to an upper limit on the inclination angle of 47 degrees.  Combining
this limit with the \omdot\ measurement and the mass function gives a
lower limit on the companion mass of 1.29 \msun.

\subsubsection{Decay of the orbit: \pbdot}\label{section:pbdot}

Another effect of the wide orbit of the system is that the decay of
the orbit due to gravitational wave damping will be very small.  
The relative motion effects, due to the proper motion and the
acceleration in the Galactic potential, will therefore completely
dominate observed changes in \pbdot (e.g. \citealt{bb96, nt95}).  Even
with the relatively low proper motion measured for this system, the
Shklovskii effect \citep{shk70} on the orbit will be
\begin{equation}
\dot P_\mathrm{b,S}= \frac{P_\mathrm{b}}{c} \frac{v_\mathrm{t}^2}{d} = 9.5\times10^{-14}
\end{equation}
where we have used the aforementioned distance of 625 pc. 

The acceleration toward the disc, though of opposite sign,
has roughly the same magnitude.  The net bias of \pbdot is probably
in the range $\Delta \dot{P}_\mathrm{b}\sim -(1-3)\times 10^{-14}$, depending
on the distance to the pulsar.  The uncertainty in this bias
is much larger than the expected $\dot{P}_\mathrm{b}$ due to general
relativity, $\dot{P}_\mathrm{b,GR}\sim-1.2\times 10^{-15}$.

Our timing solution yields an upper limit of \pbdot~$<2.4 \times 10^{-13}$. 
This is considerably larger than that expected from relativistic decay
of the orbit or from kinematic effects in the Galaxy. 
Although \pbdot may be the first other PK parameter to be measured as
the data set is extended, the combination of the three effects prevent
the use as either an independent distance estimation or the separation
of the masses of the system.

\section{Discussion}

\subsection{Masses}\label{section:masses}

\begin{figure}
  \centering
  \includegraphics[width=6.4cm, angle=270]{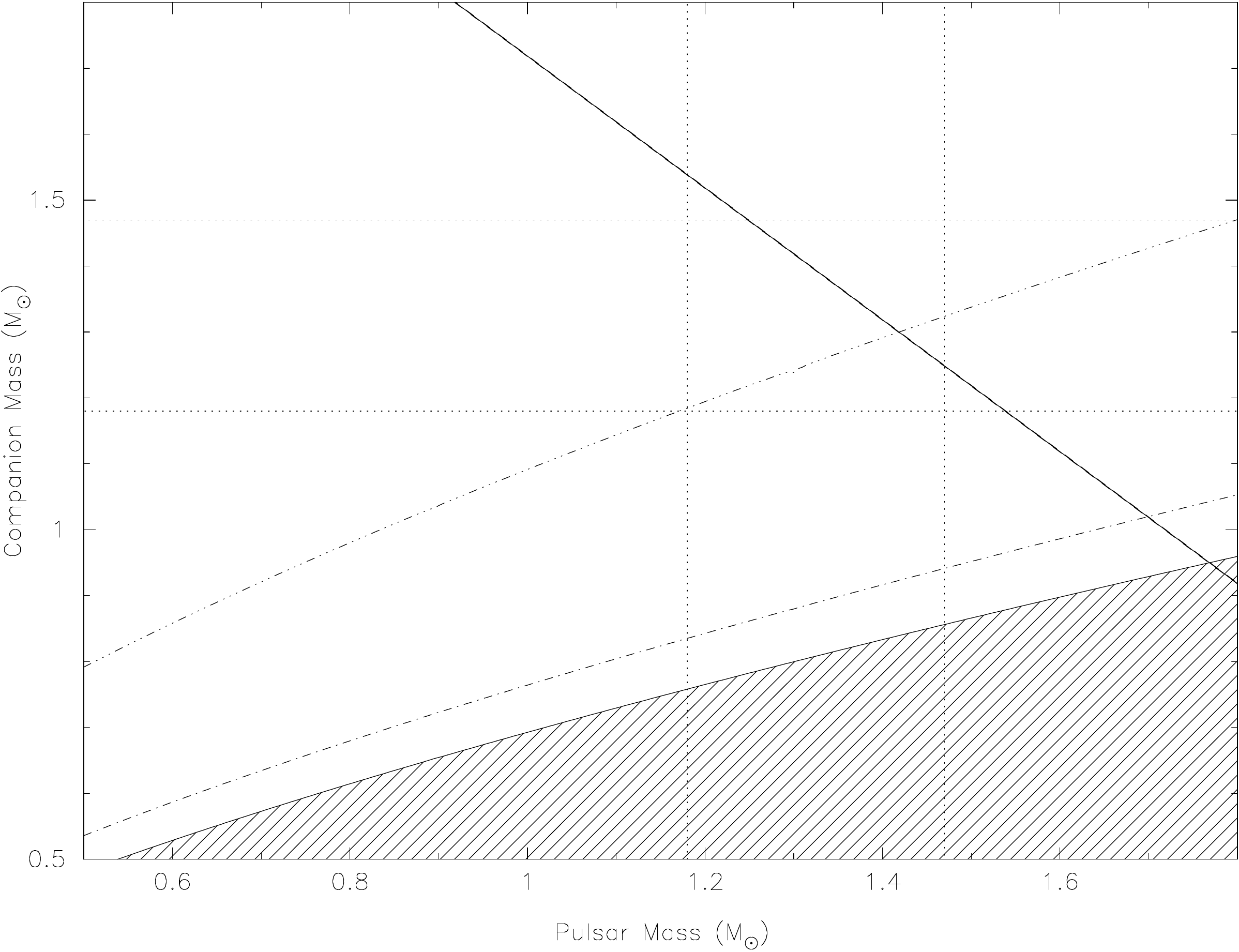}
  \caption{Mass-mass diagram for \psr and its companion. The hatched
   region is excluded for $i > 90$ degrees, the dash-dotted lines are
   indicating constraints resulting from the \xdot\ measurement, $i
   < 69$ degrees and shapiro delay limit, $i < 47$
   degrees. 
   The diagonal line constrains $\dot\omega$, with
   errors. The dotted lines for constant pulsar and companion mass
   indicate the range of neutron star masses measured in other DNSs
   \citep{tc99,lbk+04,fkl+05}.
  \label{fig:massplot}}
\end{figure}

For any given theory of gravity, measuring PK parameters will put
constraints on the masses and the inclination of the system, depending
only on the individual masses and the Keplerian parameters (e.g. \citealt{sta03}).
Figure~\ref{fig:massplot} shows all the current restrictions we were
able to derive from our timing solution (Table~\ref{tab:jwnge-2}).
The improved, highly significant \omdot\ measurement gives us a very
accurate determination of the total mass. From the non-detection of
Shapiro delay parameters it is clear that the inclination of the orbit
is quite low.  Furthermore, the $\dot{x}$ measurement presented in \S
3.3, taken at face value, confirms that the system is at low
inclination angles.

The relativistic time-dilation/gravitational redshift parameter,
$\gamma$, is only measurable when the periastron advance is large
enough to decouple its effect on the timing residuals from the
measurement of the semimajor axis and its time derivative \citep{bt75,
mt77, dt92}. Changing $\omega$ by only a few degrees will take several
hundred years, and a detection of $\gamma$ can therefore not be
expected in the near future.
Moreover, \xdot\ turns out to be highly covariant with $\gamma$
in the timing analysis.  This makes an independent measurement of
$\gamma$ impossible and obscures the interpretation of the apparent
measurement of \xdot.

To clarify the influence of all the PK perturbations on
our timing data, and to calculate accurate, refined values of the
stellar masses, we performed a comprehensive, self-consistent timing
analysis which simultaneously incorporated all relativistic and
kinematic phenomena in the timing solution.

Our primary goal was to measure or constrain the values of
$m_\mathrm{p}$ and $m_\mathrm{c}$.  This procedure also finds
constraints on the two angles which describe the orientation of the
orbit, $i$ and $\Omega$.  For convenience, we always represent the
latter as an offset from the proper motion position angle, i.e.
$\Theta_\mathrm{\mu}-\Omega$.  This angle runs between $-180^\circ$
and 180$^\circ$, while inclination runs between $0^\circ$ and
$180^\circ$.

The four degrees of freedom in this problem ($m_\mathrm{p}$,
$m_\mathrm{c}$, $i$, and $\Theta_\mathrm{\mu}-\Omega$) are reduced to
three by noting that the two masses and the inclination are related via the
Keplerian mass function equation:
\begin{equation}\label{eqn:kepler}
f\equiv\frac{(m_\mathrm{c}\sin i)^3}{(M_\mathrm{T})^2} =
\frac{x^3}{T_\mathrm{\odot}}\left(\frac{2\pi}{P_\mathrm{b}}\right)^2,
\end{equation}
where all quantities on the right side of the equation are measured to
high precision.  We analysed a grid of timing
solutions in a three-dimensional parameter space following the approach as
explained in detail in \cite{sna+02, sns+05}.
For the three variables, we used $M_\mathrm{T}$, $\cos i$, and
$\Theta_\mathrm{\mu}-\Omega$, and we assumed a uniform prior in each of
these.  For randomly oriented binary systems, $\cos i$ and
$\Theta_\mathrm{\mu}-\Omega$ each follow uniform distributions,
and since \omdot\ is well known, only a small range of $M_\mathrm{T}$
values need be considered, so a uniform prior is acceptable for this
variable as well.
For each point in our three dimensional grid, we used
Eq.~\ref{eqn:kepler} to calculate the value of $m_\mathrm{c}$
and the corresponding $m_\mathrm{p}$.
We then used the masses and orientation angles to calculate the
kinematic perturbation parameters (Eqs.~\ref{eq:xdot},~\ref{eq:omdot}) and relativistic
timing parameters according to GR (e.g. \S\,4.1 of \cite{sta03}).  

\begin{figure}
  \centering
  \includegraphics[width=7.0cm, angle=270]{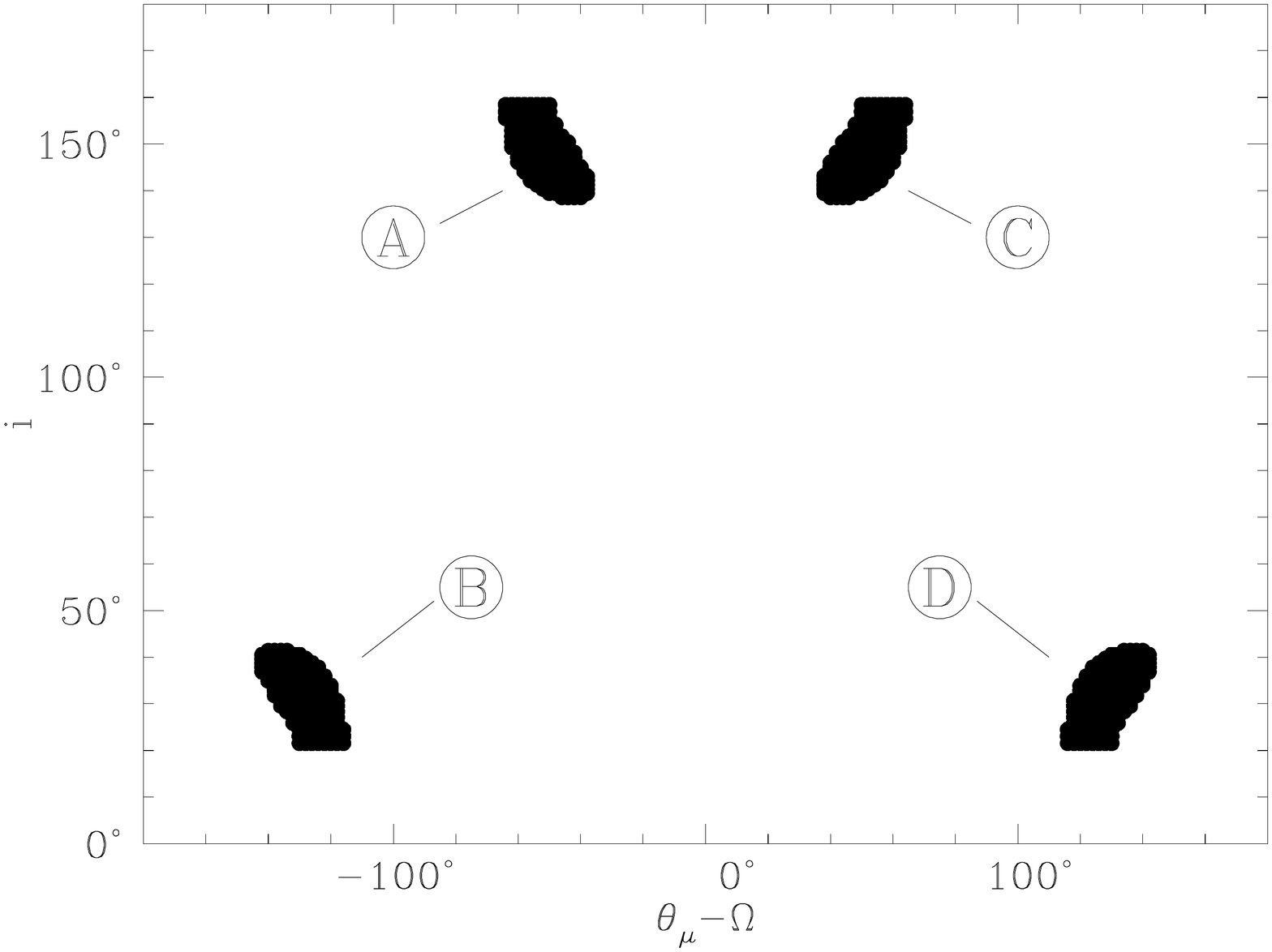}
  \caption{Allowed values for orbital orientation $i$ and
    $\theta_\mathrm{\mu}-\Omega$.  The figure shows a projection of
    the 95.4\% confidence volume of the three dimensional grid
    analysis of timing solutions onto this two dimensional space.  See
    the text for discussion.
  \label{fig:abcd}}
\end{figure}

\begin{figure}
  \centering
  \includegraphics[width=7.0cm, angle=270]{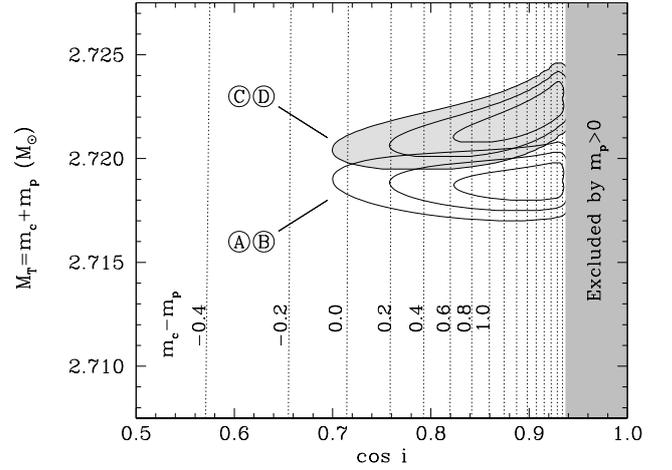}
  \caption{Allowed values for total mass and $\cos~i$.  The two sets
    of contours are 68.3\%, 95.4\%, and 99.7\% confidence limits on
    these quantities for solutions $A$ and $B$ (lower, white contours)
    and $C$ and $D$ (upper, gray contours).  Dotted lines indicate
    values of $m_\mathrm{c}-m_\mathrm{p}$, at intervals of 0.2\msun.
    The dark region on the right of the plot is excluded as it does
    not satisfy $m_\mathrm{p}>0$.
  \label{fig:chi7}}
\end{figure}

\begin{figure}
  \centering
  \includegraphics[width=10.0cm]{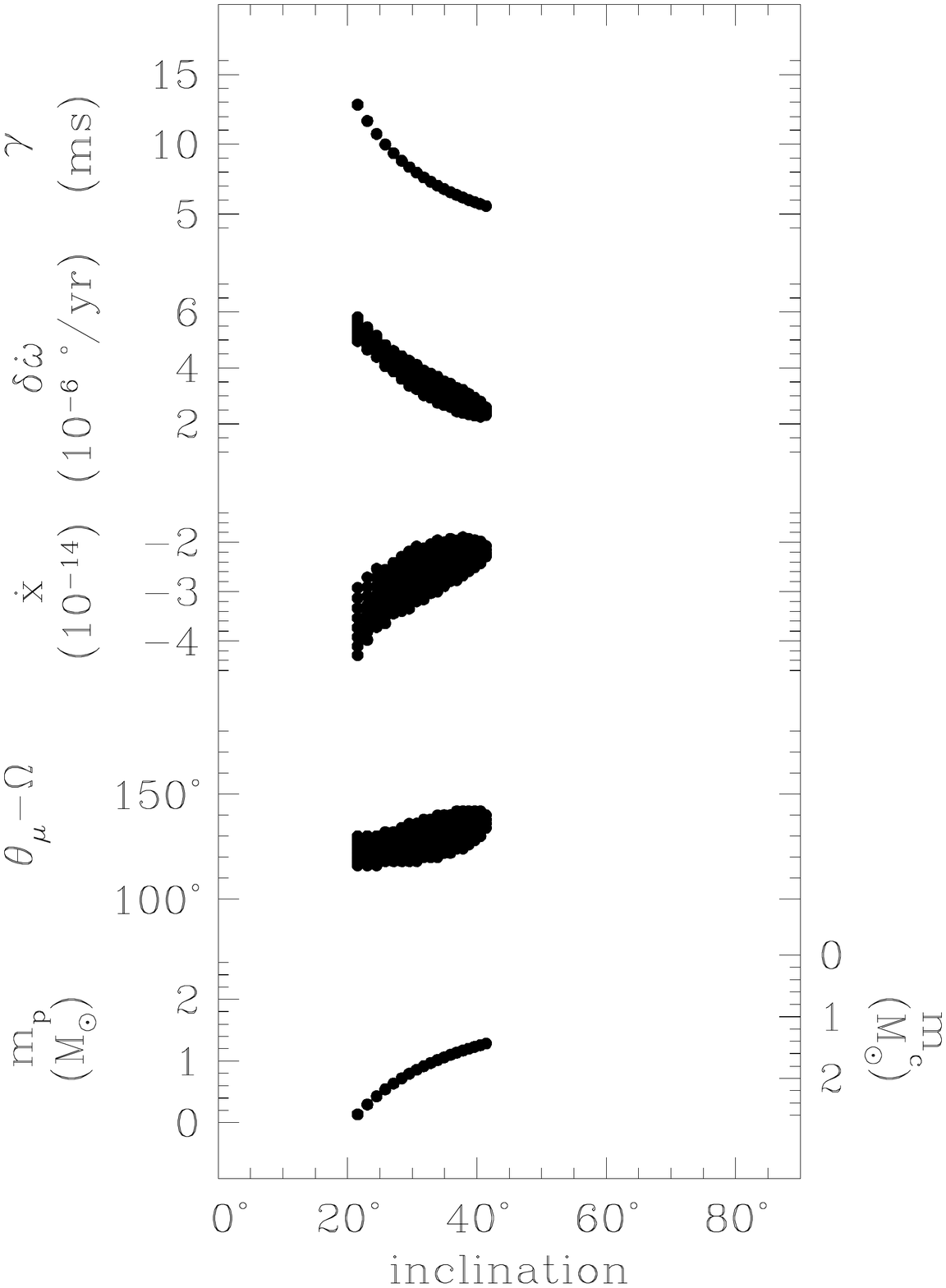}
  \caption{Values of relativistic $\gamma$, kinematic
    $\delta\dot{\omega}$, kinematic $\dot{x}$, orbital orientation
    $\theta_\mathrm{\mu}-\Omega$, and masses $m_\mathrm{p}$ and
    $m_\mathrm{c}$ allowed by the timing solution, as a function of
    inclination angle.  The figure shows a projection of the 95.4\%
    confidence volume of the three dimensional grid analysis of timing
    solutions onto each of the two dimensional spaces shown.  The
    regions corresponds to solution $D$, but other solutions give
    essentially identical results.
  \label{fig:paramplot}}
\end{figure}

The \omdot\ used in the analysis was the sum of the kinematic and
relativistic terms held fixed while all other pulsar timing parameters
(astrometric, rotational, and Keplerian orbital parameters) were free
to vary.  We recorded the $\chi^2$ of each timing solution, assigning
a probability to each grid point based on the difference between its
$\chi^2$ and the global minimum. We used this ensemble of
probabilities to calculate confidence regions for the parameters of
interest and to place limits on the stellar masses.  The results are
given in Figs. \ref{fig:abcd}, \ref{fig:chi7} and \ref{fig:paramplot}.

Figure \ref{fig:abcd} shows the orientations of the orbit allowed by the
timing solution.  The figure is a projection of the 95.4\% confidence
volume of the three dimensional analysis grid onto the two dimensional
space shown.  There are four regions allowed by the timing data.  The
Keplerian orbital parameters, combined with the
lack of Shapiro delay restrict $i$, but because the Shapiro delay
depends on $\sin~i$, the resulting constraint is degenerate: if $i$\,
is allowed, so is $180^\circ-i$.  Since there is a detectable \xdot,
Eq. \ref{eq:xdot} restricts $\Theta_\mathrm{\mu}-\Omega$ to two
possibilities (one positive and one negative value) for any given
value of $i$.  Thus there are four regions of allowed solutions,
labeled $A$, $B$, $C$ and $D$ in the figure.

Figure \ref{fig:chi7} shows 68.3\%, 95.4\%, and 99.7\% confidence limits on $\cos~i$
and $M_\mathrm{T}$ space.  For reference, the figure shows lines of
constant mass difference $m_\mathrm{c}-m_\mathrm{p}$, at intervals of
0.2\msun, as calculated from $i$ and $M_\mathrm{T}$, and it indicates
the region at low inclination angle that is excluded because $m_\mathrm{p}>0$ is not
satisfied in that region. The confidence limits were calculated by
marginalizing over the probability values for values of
$\Theta_\mathrm{\mu}-\Omega$ for a given combination of $\cos~i$ and
$M_\mathrm{T}$.

The best-fit solutions have relatively high
$m_\mathrm{c}-m_\mathrm{p}$, i.e. high companion masses and low pulsar
masses, although the confidence contours stretch to much lower values
of $m_\mathrm{c}-m_\mathrm{p}$.  The determination of $M_\mathrm{T}$
is dominated by the relativistic \omdot, but the obtained values are
perturbed by the kinematic $\delta$\omdot, given by
Eq. \ref{eq:omdot}. This perturbation splits the allowed values for
the total mass into two regions.  This can be understood by plugging
the values of $\Theta_\mathrm{\mu}-\Omega$\, from each of the four
regions of Fig.~\ref{fig:abcd} into Eq.~2, and noting that $\csc~i$ is always
positive, so that $\delta$\omdot\ is positive for solutions $A$ and
$B$ and negative for solutions $C$ and $D$.  This means that the
observed $\delta$\omdot\ has been biased towards higher values for
solutions $A$ and $B$, so that the true relativistic $\delta$\omdot\ is
lower than that calculated without the kinematic correction.  Since
$M_\mathrm{T}$ is proportional to $\delta$\omdot$^{3/2}$, this means
that the true total mass is lower for solutions $A$ and $B$.
Similarly, the true mass is higher for solutions $C$ and $D$.

Figure \ref{fig:paramplot} shows the 95.4\% confidence volume projected into several
two-dimensional parameter spaces.  For this figure, the values of $i$
and $\Theta_\mathrm{\mu}-\Omega$ correspond to solution $D$, but the
distributions of all other quantities are essentially identical for
all four solution regions. We note that, for completeness, all possible values for the
parameters are shown in Figs. \ref{fig:abcd}, \ref{fig:chi7} and
\ref{fig:paramplot} however we consider it unlikely that the pulsar
mass will be lower than $1$\msun.

We used the probabilities from the grid analysis to constrain
$m_\mathrm{p}$ and $m_\mathrm{c}$.  The results are essentially
identical for all four solution regions.
The central 95.4\% confidence intervals are
$m_\mathrm{p}=0.72^{+0.51}_{-0.58}$\,\msun and
$m_\mathrm{c}=2.00^{+0.58}_{-0.51}$\,\msun.  
If, instead of central confidence intervals, we use 
95.4\% confidence upper and lower limits for the individual
masses, these convert to $m_\mathrm{p}<1.17$\,\msun and
$m_\mathrm{c}>1.55$\,\msun. Note that the apparent discrepancy between
these numbers originates from different areas covered in the probability
distribution.
For the total mass, the 95.4\% confidence intervals are
$2.7188\pm0.0011$\,\msun for solutions $A$ and $B$, and
$2.7217\pm0.0018$\,\msun for solutions $C$ and $D$.  However, allowing any
of the four solutions yields the range $2.720_{-0.002}^{+0.003}$\msun.
This last number is the most accurate value for $M_\mathrm{T}$, as it
reflects the uncertainty as to which orientation of the orbit is
correct.

\subsection{Evolution}

The system is valuable for studies of DNS evolutionary scenarios.  It
has been believed for a long time that supernova explosions result in
high kick velocities on the remaining neutron star
\citep{bai89}. Recently, it has been argued that different
evolutionary scenarios can be possible in binary systems with
particular mass and chemical properties \citep{plp+04,heu07}. The
so-called electron-capture collapse of an O-Ne-Mg core of a helium
star leads to a {\it fast} supernova explosion which is believed to be
symmetric and therefore does not result in a kick of the second-born
neutron star.  For most DNSs in which proper motion measurements have
been possible, a low space velocity has been derived.
Our measurement of the proper motion of \psr, implying a velocity of
about 25 km\,s$^{-1}$, is consistent with the evolutionary scenario
mentioned above, and another piece of evidence that not all pulsars
receive a large kick at birth.

Van den Heuvel (2007) argues that the low velocity of DNS systems may
be correlated with the masses of the second-formed neutron stars in DNSs
being somewhat lower than normal: $1.25(6)$\msun.  Our 95.4\%
limit\footnote{The 99.7\% limit is $m_\mathrm{c} > 1.39$\msun} of
$m_\mathrm{c} > 1.55$\msun appears to contradict this correlation. 

In case of a symmetric supernova the mass lost during the explosion
can be calculated by the expression $e=\Delta M_\mathrm{SN}/M_\mathrm{T}$
\citep{bv91}, where $e$ is the eccentricity and $M_\mathrm{T}$ the total mass
after the supernova explosion. For \psr\ this results in $\Delta
M_\mathrm{SN}=0.68$\msun.  If we use our 95.4\% lower limit on the
companion of 1.55~\msun this leads to a core-progenitor mass of
2.23\msun, which is considered to be in the lower range for helium
cores \citep{dp03b}, and the progenitor star must have had a mass of
about $9$\msun. As these stars are believed to produce degenerate
O-Ne-Mg cores \citep{plp+04,phlh08}, this is another indication that
the companion neutron star must have formed in an electron-capture
core collapse. The somewhat higher companion mass can possibly be
explained by assuming extra fall-back during the supernova event of
$\sim0.2$\msun (e.g. \citealt{fk01,fry07}). To allow for this amount
of fallback the supernova explosion has probably been very weak.
An electron-capture core collapse is the most probable explanation for
the parameters of this system. 
However, constraining the masses even better would be very interesting.

\subsection{Search for the companion}

Apart from the original discovery papers \cite{nst96,snt97}, there is
no report of searches for pulsations of the companion of \psr. The
Green Bank Northern Sky Survey, optimized to find
millisecond pulsars, had a flux-density limit at 370 MHz of 8 mJy for
slow pulsars (with periods above 20 ms). 

The very high characteristic age and the low magnetic field
imply that \psr is the first-born and recycled neutron
star in the system. 
The (unseen) second neutron star will be the young object, with a long
spin period, which has probably already slowed down significantly and
possibly passed the death-line to become undetectable as a radio
pulsar. Moreover, if the second neutron star is still active as a
radio pulsar, it has a reasonable chance of being beamed away from
us. In this case, taking into account that the precession timescale of
the spin axis is very long, it is not expected that the companion, if
active as a pulsar, will rotate into view on a short timescale.

However to be certain that no weak pulsed signal from the second
neutron star has been missed, we have searched several WSRT
observations of 30 minutes at 840, 1380 and 2300 MHz for pulsations of
the companion of \psr.  
We performed an acceleration search on the data
although, because of the wide orbit, the expected smearing due
to the orbital motion of the companion is likely to be negligible.

In the only double pulsar system known so far, PSR~J0737$-$3039A/B,
the second pulsar is only visible for a small part of each orbit
\citep{lbk+04}.  To take this possibility into account, and assuming
the second pulsar would be visible in an equal fraction of the orbit
as PSR~J0737$-$3039B, we have searched another set of observations
spread equally across the full orbital phase range. In all searches no
evidence of pulsations of a companion were found to the limits of 1.1,
0.25 and 0.5 mJy at 840, 1380 and 2300 MHz respectively, which means
that if the companion is active as a pulsar, and beamed towards us, it
must be less luminous than 0.098 mJy~kpc$^2$ at 1380 MHz.

\section*{Conclusions}

Using more than 12 years of timing observations obtained by 5
telescopes, we have improved the timing solution of \psr. We 
accurately determined the total mass of the system to be
$M_\mathrm{T}=2.7183(7)$\msun and the proper motion $\mu_\mathrm{T} =
8.55(4)$ mas\,yr$^{-1}$. From our timing solution we were able to
set constraints on the individual masses of the system:
$m_\mathrm{p}=0.72 _{-0.58}^{+0.51}$\msun and $m_\mathrm{c}=2.00
_{-0.51}^{+0.58}$\msun (95.4\%), and the inclination of the orbit:
$i<47$ degrees (99\%).

\subsection*{Future prospects}

\begin{figure}
  \centering
  \includegraphics[width=6.5cm, angle=270]{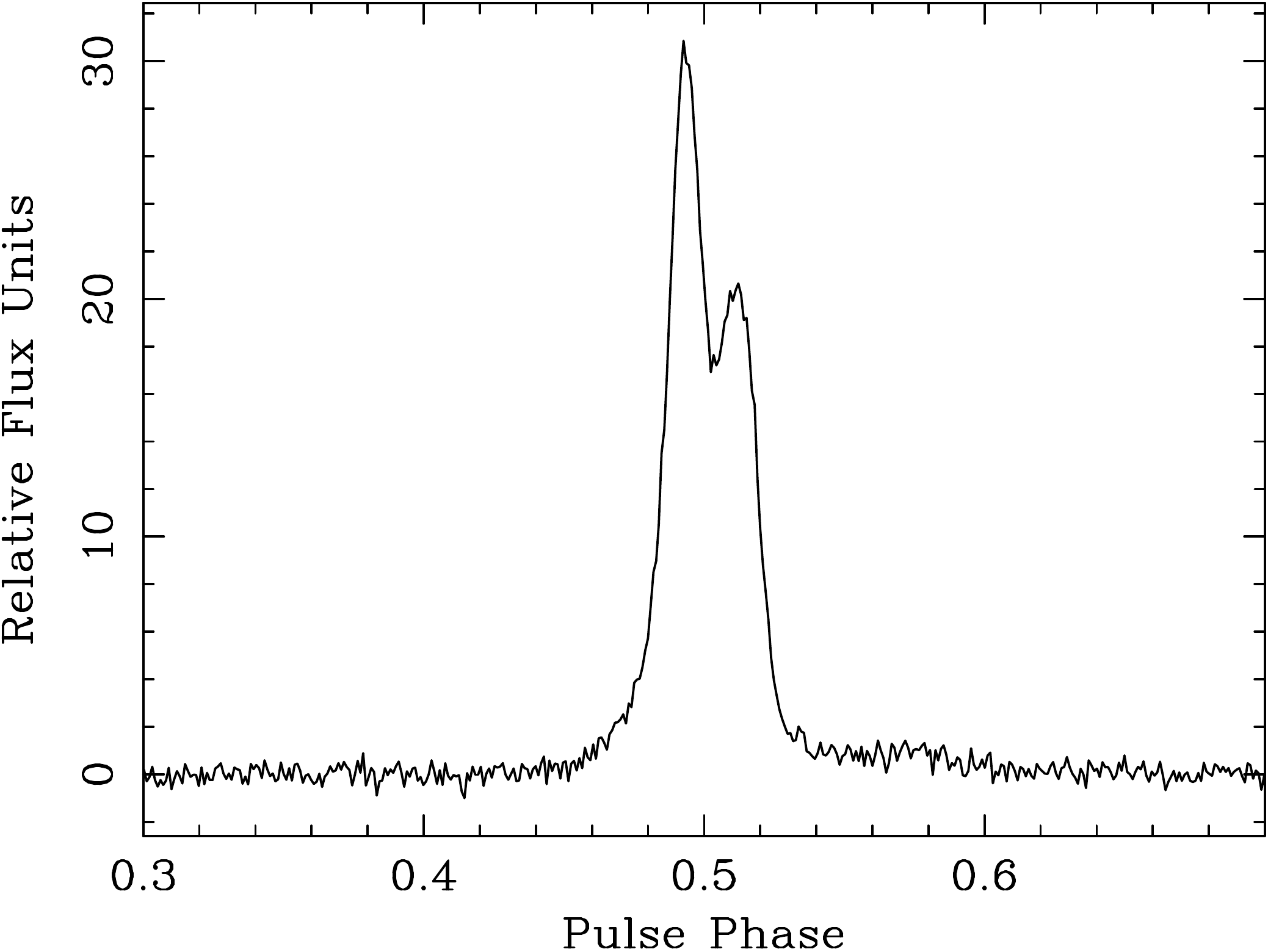}
  \caption{A pulse profile obtained using PuMa II at the WSRT. The
data were obtained in a 30 minute observation at 1380 MHz with a bandwidth
of 160 MHz and were analysed using a coherent filterbank with 512
frequency channels and a time resolution of 6.4 $\mu$s. A similar
pulse phase range is shown as in Fig. \ref{fig:stds} and the improved resolution
shows somewhat more detail in the pulse profile.
  \label{fig:puma2}}
\end{figure}

As discussed above, at the present timing precision it will not be
possible to detect any further PK parameters in this system on any
reasonable timescale. Recently we have begun observing the system
using the new PuMa II coherent dedispersion backend at the WSRT
\citep{ksv08}. These observations use a total of 160 MHz of bandwidth
and thus should have at least a factor $\sqrt{2}$ improved SNR.  When
combined with the somewhat narrower profile obtained using coherent
dedispersion (see Fig.\ref{fig:puma2}), this should lead to improved arrival time
measurements. At present we have insufficient data to check on the
long term timing precision, however the error in the individual
arrival times are at least a factor two better than the previous
data. These improvements might allow us to put stronger limits on the PK
parameters in this system.

\section*{Acknowledgements}

The Westerbork Synthesis Radio Telescope is operated by ASTRON
(Netherlands Foundation for Research in Astronomy) with support from
the Netherlands Foundation for Scientific Research NWO.  
The Nan\c cay radio Observatory is operated by the Paris Observatory,
associated to the French Centre National de la Recherche Scientifique
(CNRS). The Nan\c cay Observatory also gratefully acknowledges the
financial support of the Region Centre in France.
DJN is supported by NSF grant AST-0647820 to Bryn Mawr College.
The 140 Foot Telescope and the Green Bank Telescope are facilities of
the National Radio Astronomy Observatory, operated by Associated
Universities, Inc., under a cooperative agreement with the NSF.
MBP acknowledges support from the Science and Technology Facilities
Council (STFC). 

We would like to thank Jason Hessels for help with the search for the
companion, Kosmas Lazaridis for help with Effelsberg data analysis,
and Ed van den Heuvel for helpful discussions on the evolution of
DNSs.

\end{document}